\documentclass[preprint,aps,prl]{revtex4}
\usepackage[dvipdf]{graphicx}
\usepackage{dcolumn}
\usepackage{bm}
\usepackage{amssymb}
\usepackage{xcolor}

\begin{document}
\title{Gapless ground state in the archetypal quantum kagome antiferromagnet ZnCu$_3$(OH)$_6$Cl$_2$}
\author{P. Khuntia$^{1,2}$, M. Velazquez$^{3,4}$, Q. Barth\'elemy$^1$, F. Bert$^1$, E. Kermarrec$^1$, A. Legros$^1$, B. Bernu$^5$, L. Messio$^{5,6}$,  A. Zorko$^7$, P. Mendels$^1$}
\affiliation{$^1$ Laboratoire de Physique des Solides, Universit\'e Paris-Sud 11, Universit\'e Paris-Saclay, UMR CNRS 8502, 91405 Orsay, France.\\
$^2$ Department of Physics, Indian Institute of Technology Madras, Chennai-60036, India.\\
$^3$ CNRS, Universit\'e de Bordeaux, Bordeaux INP, ICMCB, UMR 5026, 33608 Pessac, France.\\
$^4$ Univ. Grenoble Alpes, CNRS, Grenoble INP, SIMAP, 38000 Grenoble, France.\\
$^5$ Laboratoire de Physique Th\'eorique de la Mati\`ere Condens{\'e}e, LPTMC, Sorbonne Universit\'e, CNRS,  F-75005 Paris, France.\\
$^6$ Institut Universitaire de France (IUF), F-75005 Paris, France.\\
$^7$ Jo\v{z}ef Stefan Institute, Jamova c.~39, SI-1000 Ljubljana, Slovenia.}

\maketitle

{\bf Spin liquids are exotic phases of quantum matter challenging Landau's paradigm of symmetry-breaking phase transitions. Despite strong exchange interactions, spins do not order or freeze down to zero temperature. While well-established for 1D quantum antiferromagnets, in higher dimension where quantum fluctuations are less acute, realizing and understanding such states represent major issues, both theoretical and experimental. In this respect the simplest nearest-neighbor Heisenberg antiferromagnet Hamiltonian on the highly frustrated kagome lattice has proven to be a fascinating and inspiring model. The exact nature of its ground state remains elusive and the existence of a spin-gap is the first key-issue to be addressed to discriminate between the various classes of proposed spin liquids. Here, through low-temperature Nuclear Magnetic Resonance (NMR) contrast experiments on high quality single crystals, we single out the kagome susceptibility and the corresponding dynamics in the kagome archetype, the mineral herbertsmithite, ZnCu$_3$(OH)$_6$Cl$_2$. We firmly conclude that this material does not harbor any spin-gap, which restores a convergence with recent numerical results promoting a gapless Dirac spin liquid as the ground state of the Heisenberg kagome antiferromagnet.}

In the context of magnetism, Quantum Spin Liquids (QSL) appear exotic as they harbor a quantum-entangled ground state with no symmetry breaking  and unconventional excitations, e.g. fractional ones such as spinons or Majorana fermions~\cite{Knolle}. A large variety of such QSL could be classified theoretically~\cite{Wen,Savary}. In parallel, the pool of experimental candidates built from a regular assembly of frustrated motifs and, more recently, bond-dependent interactions in the Kitaev case has been steadily increasing over the past 20 years encompassing Mott-Hubbard organic materials and deep Mott insulators in various two and three dimensional geometries~\cite{Lacroix-Springer, Balents, Mendels-JPSJ, Mendels-academy, Zhou-KanodaRMP}. Among them, the kagome lattice is a prominent 2D example where frustrated triangles only share corners. This reduced lattice connectivity, the frustration generated by nearest neighbor Heisenberg antiferromagnetic interactions, and the quantum character of $S=1/2$ spins indeed conspire to stabilize such a highly entangled QSL state in  herbertsmithite~\cite{Shores}, ZnCu$_3$(OH)$_6$Cl$_2$ (fig.~1). It emerged as the first kagome based antiferromagnet featuring a perfect equilateral triangular geometry, dominant nearest neighbor Heisenberg antiferromagnetic interactions, $J\sim180-190$~K, and not showing any ordering~\cite{Mendels-musr} but rather an unconventional excitations continuum~\cite{Han-neutrons}. It is recognized as the kagome spin liquid candidate approaching so far the best the ideal model of an Heisenberg AntiFerromagnetic Hamiltonian on the Kagome lattice (KHAF)~\cite{Norman}, now available in single crystal form. Its discovery triggered a burst of experimental activity, on one hand with the materials search for Cu$^{2+}$-based $S=1/2$ variants  obtained either from a change of anions~\cite{Barlowite, BarlowiteNMR, Brochantite} or replacement of Zn$^{2+}$ cations~\cite{PuphalGa, PuphalY, ChineseY, BartelemyY} and Li$^+$ intercalation in view of charge doping~\cite{McQueen}, and on the other hand with V$^{4+}$-based $S=1/2$~\cite{VOF,Orain} QSL candidates.

On the theory side which was also revived by the discovery of herbertsmithite, the issue of the ground state of the KHAF, despite its apparently simple form, has been lacking for long a definite conclusion. The main reason lies in the existence of a proliferation of states close-by in energy which have led to theoretical proposals spanning from valence bond crystals~\cite{Singh} - made of local spin dimers- to spin liquids, either gapped like the resonating valence bond state, or not, like the Dirac U(1) spin liquid~\cite{Ran}. Each novel round in the debate has resulted from challenging developments of the numerical techniques~\cite{Lhuillier,Yan,Depenbrock,Iqbal,Lauchli,Hotta,He,Normand}. 
While Density Matrix Renomalization Group (DMRG) first concluded that a gapped RVB state could be stabilized~\cite{Yan,Depenbrock}, variational Monte-Carlo (VMC)~\cite{Iqbal} methods, exact diagonalizations (ED) for a 48 spins cluster ~\cite{Lauchli}, "grand canonical analysis"~\cite{Hotta} and further works using DMRG suggested that this is in fact not the case, recently pointing to a gapless Dirac state~\cite{He} which is also found using Tensor Network States (TNS)~\cite{Normand}. In this context, the issue of the gap in the KHAF model which we address here experimentally, is a corner-stone in the investigations of this model.

$^{17}$O NMR, as a local technique, can resolve that issue and sets in as both a unique and very sensitive probe of the copper-oxygen kagome planes in the model compound herbertsmithite. Indeed, oxygens are on the exchange path connecting kagome spins (fig.~1b) and, as we show here, are much less coupled to inter-plane Cu's (ratio of the coupling constants less than 1:23) which appear as the main defects in herbertsmithite. 
These defects mask the physics from the kagome spins through dominant Schottky~\cite{deVries, Han} and Curie-like~\cite{Bert, Han-susc} contributions in low-$T$ specific heat and susceptibility experiments, respectively. Initially used on powders, $^{17}$O NMR gave the first hint about the low-$T$ kagome susceptibility from shift measurements and spin dynamics from spin-lattice relaxation time $T_1$ measurements and also revealed the existence of defect sites~\cite{Olariu}. Based on a more accurate study once single crystals became available, Fu et al. argued in favor of a gapped state in herbertsmithite with a gap closing under an applied field~\cite{Fu}. In this paper, we carry out a still more refined $^{17}$O NMR study on single crystal, relying both upon contrast experiments similar to those used in Magnetic Resonance Imaging (MRI) and upon $T_1$ spin-lattice relaxation time measurements. This enables us to isolate the susceptibility of the kagome spins far from defects, and to study their dynamics. Based on this, we come to the firm conclusion that the ground state in herbertsmithite is gapless.
Concomitantly, we single out a very specific signature induced by the defects due to Cu/Zn substitution and show that a screening occurs around these defects, which sets in as a landmark of the spin liquid state.

Our $^{17}$O NMR data were taken on an enriched crystal, grown following the same route as described in~\cite{Han-susc} and used in~\cite{Fu}, leading to high quality crystals (see SI.~1).  An NMR spectrum with 5 equidistant lines is in general expected for a given oxygen environment and both contributions, far and near to defects can be resolved at high $T$, e.g. for a field applied along $c$ (SI.~2).
The spectral resolution for such a field direction is progressively lost at low-$T$, the focus of this study, due to spectral broadening. It is much improved by combining a field applied along $a^*$ as established in~\cite{Fu} and contrast methods introduced here for this purpose.
One main quintet of lines (M) emerges which reveals the physics of the kagome spins far from defects. By tracking the shift of that line, one can extract the variation of the intrinsic kagome susceptibility and probe the dynamics through relaxation time measurements.

In fig.~2a, spectra taken between 100 and 20~K are displayed. While the main line shifts with lowering temperatures towards the zero-susceptibility reference as often observed for correlated antiferromagnets, a progressive broadening is observed and, noteworthy, a new series of less intense lines, ascribed to oxygens close to defects, emerges which becomes much better resolved than the main line. Fig.~2b illustrates at 10~K the results of the two contrast methods used here, which are crucial to extract the intrinsic shift far from defects, {\it i.e.} associated with the kagome physics in isolation. The latter contribution has shorter transverse ($T_2$) and longitudinal nuclear relaxation times ($T_1$) so that (i) by varying the time between pulses in a spin echo experiment ($T_2$ contrast), one can isolate the contribution from O sites nearby a defect at long times (see also SI.~3); (ii) by using a saturation-recovery sequence ($T_1$ contrast), one can severely decrease the defect contribution and get the unbiased defect-free $^{17}$O spectra revealing the intrinsic kagome susceptibility. Fig.~2b shows clearly the difference between the defect free (M) and the impacted (M+D) spectra. In fig.~2c, a series of defect-free spectra obtained with method (ii) are displayed for different temperatures below 10~K.

While the substantial broadening of the lines yields large error bars on the {\it absolute} determination of the shift at each $T$, we discovered that scaling all the contrast spectra taken around 6.7~T by their width and using a simple shift translation brings them into a remarkable coincidence (fig.~3a). This by far improves the accuracy on our determination of the {\it relative} variation of both the width and the shift (fig.~3b-c). Additionally, at a given $T$, using the same width scaling procedure for various applied fields (fig.~3a) shows that the shift does not depend on the field in the range 2.6 - 6.7 T (fig.~3b-c and SI.~3). Fig.~3c clearly shows the difference of the $^{17}$O shifts for defect-free spectra as compared to raw spectra. While the latter mixes the main and the defect contributions and thus artificially displaces the maximum of the line towards zero shift, which can be mistakenly interpreted as the evidence for a gap, our analysis singles out the pure kagome contribution which turns out not to be gapped.

In more detail, the shift $K_{\rm M}$ of the main line combines two contributions, one, $K_{\rm spin}$, proportional to the local spin susceptibility and the other one, the chemical shift, $K_{\rm chem}$, which is independent of temperature, $K_{\rm M} = K_{\rm spin} + K_{\rm chem}$. The linear  $T\rightarrow 0$ extrapolation of $K_{\rm M}$ either yields a small finite or zero value of the spin shift at $T=0$, depending on the value of K$_{\rm chem}$, which is hard to estimate accurately in the case of herbertsmithite as discussed in SI.~4. Using the hyperfine coupling constant and the values of K$_{\rm chem}$ determined in SI.~4, we extract the kagome susceptibility which clearly does not show an activated behavior as suggested earlier~\cite{Fu} but rather behaves linearly in $T$.

In order to assess the gapless character of the ground state even more firmly, we also performed spin-lattice relaxation measurements which probe the imaginary part of
the dynamic spin susceptibility $\chi "(\omega)$ at very low frequencies, a stringent probe of the existence of a gap. We took our data on a part of the raw spectra where
defect lines marginally contribute (fig.~2, SI.~6).
We find a slightly sub-linear  $T^{0.84(3)}$ variation of $1/T_1$ as shown in fig.~3d. This unambiguously rules out an activation law and definitively settles a gapless ground state from the dynamical point of view.


Our results for the NMR shift, corresponding to the kagome susceptibility, can be very satisfactorily reproduced by an extension to the gapless case of the microcanonical approach developed in~\cite{BernuLhuillier}. This relies on the series expansion of the entropy, once the $T\rightarrow 0$ asymptotic variation of the specific heat $C$ has been assumed.
In the gapless case, $C \sim T^\alpha$, one can then derive the susceptibility after fixing the parameters $\chi(T=0)$ and $\alpha$ . Fixing $\chi(T=0)$ within the limits
consistent with our data and assuming either a linear or a quadratic in $T$ specific heat, these series 
account reasonably well for the shift data in the entire $T$ range with $J$ values between 180 and 190~K (see SI.~5). These values of $J$ are consistent with those derived from the high-temperature series expansion analysis of the macroscopic susceptibility, where the uncertainty is linked with the amount of Cu defects obtained in the fits (see SI.~4). The best low-$T$ fit (fig.~3b) is obtained  under the assumption of a linear asymptotic behavior of the specific heat. 

Turning now to the various numerical methods, VMC, DMRG and TNS developed to approach the ground state of the KHAF, all seem to converge towards a Dirac spin liquid ground state. In this case, one expects in zero-field~\cite{Ran} (i) a $T$-linear variation of the susceptibility; (ii) a $1/T_1$ variation algebraic in $T$, with an exponent which could be consistent with our data; and (iii) a specific heat which would vary like $T^2$ although, as already explained, the latter is hard to measure due to the quasi-free Cu on the Zn site. In our experiments, we reach at low $T$ a regime $\mu_B B > k_B T$ where spinon pockets develop at the Fermi level at the expense of the Dirac nodes~\cite{Ran}. A low-T crossover is then expected, with a finite $\chi(T\rightarrow 0)$ susceptibility proportional to the field and a linear rather than a quadratic in $T$ specific heat when $\mu_B B \gg k_B T$. By developing calculations in line with this Dirac scenario, we can fit well our data below 15~K 
with only one adjustable parameter, the Fermi velocity which is defined as the ratio of energy and momentum for low-energy excitations, predicted to be $v_F\cong4.8\times10^3$~m/s~\cite{Ran}. The value of $v_F$ that we find from a fit up to the limit where a cone picture might apply, $T<15$~K ~\cite{Ran}, is about a factor 1.2 larger, only. 
This is a reasonable agreement, given the assumptions made to derive that theoretical value~\cite{Ran}. Our simulation then explains quite well the $T\rightarrow 0$ extrapolation of the  susceptibility to finite values consistent with our data, within error bars, for $B$ varying between 2.6 and 6.7~T and $T\geq 1.3$~K.  Also, we note that the better fit with series obtained assuming a linear rather than quadratic in $T$ specific heat in the $T\rightarrow 0$ limit- which implies $\mu_B B > k_B T$- is in-line with this scenario. Although our data therefore hint at the existence of a Dirac spin liquid in herbertsmithite, further theoretical considerations are needed since these models might be unstable under a magnetic field or DM anisotropy, e.g. as seems to apply to the U(1) Dirac model developed in~\cite{Ran}.

Alternatively, in a scenario where the susceptibility at $T=0$ is finite, whatever the field, a Fermi sea of spinons could be invoked in the spirit of the models which have been proposed for the QSL state found in two organic triangular antiferromagnets~\cite{Zhou-KanodaRMP}. There, while the $T \rightarrow 0$ macroscopic susceptibility remains finite
, $T_{1}^{-1}$ changes smoothly from a moderate power law in $T$ or $T^{1/2}$ to $1/T^{1.5-2}$ at very low temperatures ($T/J<\;10^{-3}$). We note however that the $T=0$ susceptibility for these compounds is rather large when compared to the maximum susceptibility. 
This is much different from what we observe in herbertsmithite where $\chi(T=0)$ obtained from a linear extrapolation hardly represents 7\% of the maximum of the susceptibility found for $T\sim J/3$.

A random singlet state induced by defect-generated disorder has also been proposed as a possibility to explain the gapless character of the herbertsmithite ground state~\cite{Kawamura}. This requires a minimal distribution of exchange interactions, $\Delta J /J\sim 0.4$. Since disorder will affect concomitantly the overlap of copper and oxygen orbitals  responsible for both the hyperfine coupling, $A$, and for the exchange interaction, $J$, one can deduce an upper bound of 3.2~\% for $\Delta J /J$ by considering the high-$T$ linewidth to shift ratio of 3.6\% and attributing its origin equally to a distribution of $J$ and $A$ only. This is one order of magnitude smaller than required, so this scenario can be safely discarded. $\Delta J /J$ is also much too small to induce a finite density of states at zero energy which could have been invoked to explain a finite susceptibility at $T=0$ and a gapless $T_1$~\cite{graphene}.

Although our results for herbertsmithite comply with the recent convergence of models toward a Dirac spin liquid, we note, for completeness, that (i) very recent results from series expansion~\cite{Hotta} and exact diagonalizations~\cite{Schnack} point to a maximum of the susceptibility at $T(\chi_{\rm max})$ of the order of $0.1-0.15~J$, significantly lower than our value $\sim 0.3\;J$. If confirmed, these results would signal that herbertsmithite is possibly not as close to the ideal case as one could believe and, as a counterpart, the behavior of the susceptibility is dramatically sensitive to other details in the Hamiltonian. In that case one could expect the ground state to be markedly influenced by varying defect rates or by applying pressure which could vary the scheme of interactions. This draws perspectives for future investigations of herbertsmithite; (ii) recent TNS calculations including Dzyaloshinsky-Moriya (DM) anisotropy point to a transition to an ordered ground state for $D/J>0.02$~\cite{Normand-bis} smaller than the estimated range $0.044 <D/J<0.08$ given by the analysis of ESR data, in a framework where the ESR linewidth was solely attributed to DM anisotropy~\cite{Zorko-ESR1, ElShawish}. This calls for a more refined analysis, using a combination of exchange and DM anisotropies as was done in vesignieite~\cite{Zorko-Vesi} and briefly discussed in SI.~4; (iii) Our finding of a quasi T-linear  1/T$_1$ dependence calls for further theoretical efforts to compute this quantity as it may help to discriminate between the various gapless models proposed so far.

In the vein of defects physics in herbertsmithite, some precious piece of information can be retrieved from our spectral study. Indeed, as pointed above, the effect of defects on NMR spectra can be singled out either through contrast experiments revealing specific resolved spectra (D) for their neighbouring oxygens or through the increase of the linewidth of the (M) line at low-$T$, revealing an inhomogeneous susceptibility for further sites. Such signatures share a commonality with low-dimensional $S=1/2$ antiferromagnetically correlated materials such as spin chains or High-$T_c$ cuprates~\cite{Tedoldi, Alloul-review} where a staggered response develops around a spin defect, e.g. spinless or $S=1$. However for the latter, a similar broadening occurs for sites close to and far from defects. By contrast, here in
herbertsmithite, no broadening affects the (D) spectra, a striking result in-line with the ESR signature of decoupled defects~\cite{Zorko-ESR2}.

Pushing further the analysis, we note that at least 6 lines from the (D) spectra are resolved (fig.~4a), that is more than the 5 lines expected for a single defect site, revealing two types of oxygen sites close to defects whose shift variations with temperature differ considerably, as shown in fig.~4c. One site (D$_1$) displays a large negative shift which saturates at low-$T$ under the applied fields of a few Teslas (fig.~4c-d), pointing to a negative hyperfine coupling to an inter-layer Cu moment.
The well-defined narrow lines point to an insensitivity to further defects at variance with the main line. This is an intriguing finding since on the other hand the similar types of $T$-variations and low-$T$ saturations for both the width of the main line and the shift of the D$_1$ line suggest   a common origin, namely the interlayer Cu's (fig.~4 and SI.~7). The other site (D$_2$), on the contrary displays no shift and therefore reveals surrounding Cu's harboring a zero susceptibility at low-$T$.

The possible Cu environments of an oxygen site are sketched in fig.~4b~: (I) straightforwardly corresponds to "ideal" oxygen sites far from defects corresponding to the (M) line; (II) corresponds to sites which could be at the origin of both the (M) linewidth and the (D$_1$) line despite the absence of broadening. Why such a weakly coupled out-of-plane Cu located on a Zn site yields a staggered response is certainly associated with a Jahn-Teller driven distortion leading to a displacement of the six adjacent oxygens.
 The reshaped triangle then plays the role of an in-plane magnetic defect which could generate the staggered magnetic response. The environment (III) has remained more controversial as it features a Zn substituting an in-plane Cu which was advocated against in~\cite{McQueen2} at the limit of consistency between the characterization techniques. However, the arguments developed at length in SI.~8 based on the high-$T$ shifts analysis clearly support the existence of configuration (III). Moreover, it provides a natural explanation for the (D$_2$) line through a release of frustration on the two Cu's adjacent of such a configuration (III) which are then prone to dimerize into a singlet state and generate a staggered response as shown in~\cite{Rousochatzakis}.


In conclusion, in herbertsmithite, quoted as archetypal in the landscape of spin liquids as it maintains its spin liquid state down to $T\rightarrow 0$ with a model Hamiltonian dominated by {\it n.n.} interactions, we firmly settle major evidences of a gapless ground state through the $T$-dependence of the shift and of the relaxation rate. Our results pave two new research avenues to help improving our understanding of the KHAF. (i) The resolution of defect sites clearly offers a handle to probe the physics through a new prism~\cite{Kondo-Brochantite}. This "perturb to reveal" method appears as a unique tool to investigate the spin liquid nature of herbertsmithite and of other candidates. While this offers novel issues to be addressed by theoretical models, on the materials side, one is now facing the challenge of better controlling the concentration of defects in herbertsmithite or its variants. (ii) Other issues have to be addressed in order to refine the comparison of the ultimate $T\rightarrow 0$ measurements with theories which have to include calculations of $T_1$ and deviations to the Hamiltonian such as the small exchange anisotropy and the DM anisotropy which is unavoidable in as low-symmetry structures as the kagome network.

\clearpage
\begin{center}
\includegraphics[width=1\textwidth]{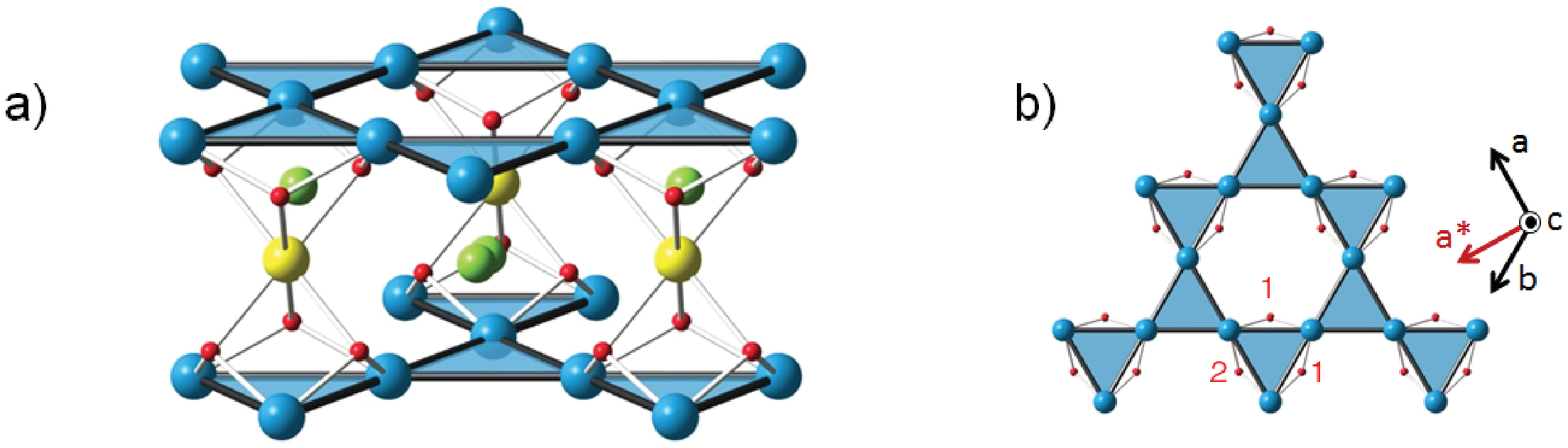}
\end{center}
Fig.~1:~{\bf Herbertsmithite structure}\\
(a) Side view of Herbertsmithite, ZnCu$_3$(OH)$_6$Cl$_2$. It consists of geometrically perfect kagome layers where Cu$^{2+}$ ions interact through an antiferromagnetic superexchange interaction, $J\sim 180$~K, mediated by bridging O$^{2-}$ ions. Cu's are represented in blue, Zn's in yellow, Cl's in green and oxygens in red. Ideally, the interstitial layer should  be filled with diamagnetic Zn$^{2+}$ ions, resulting then in quasi-
decoupled kagome layers.\\
(b) Top view of the Cu kagome layer emphasizing the ideal position of oxygens to probe the physics of the kagome S=1/2 Cu$^{2+}$ layers through Nuclear Magnetic Resonance (NMR). In our NMR experiments, the field is applied along the reciprocal-axis $a^*$. Though all the O occupy the same crystallographic site, in the latter configuration the quadrupolar hamiltonian, which is sensitive to the field versus crystal axis orientations, is different for oxygen sites 1 and 2. This yields distinct NMR spectra as clearly shown in fig.2(a) for $T\sim 60$~K.
\clearpage
\begin{center}
\includegraphics[width=0.9\textwidth]{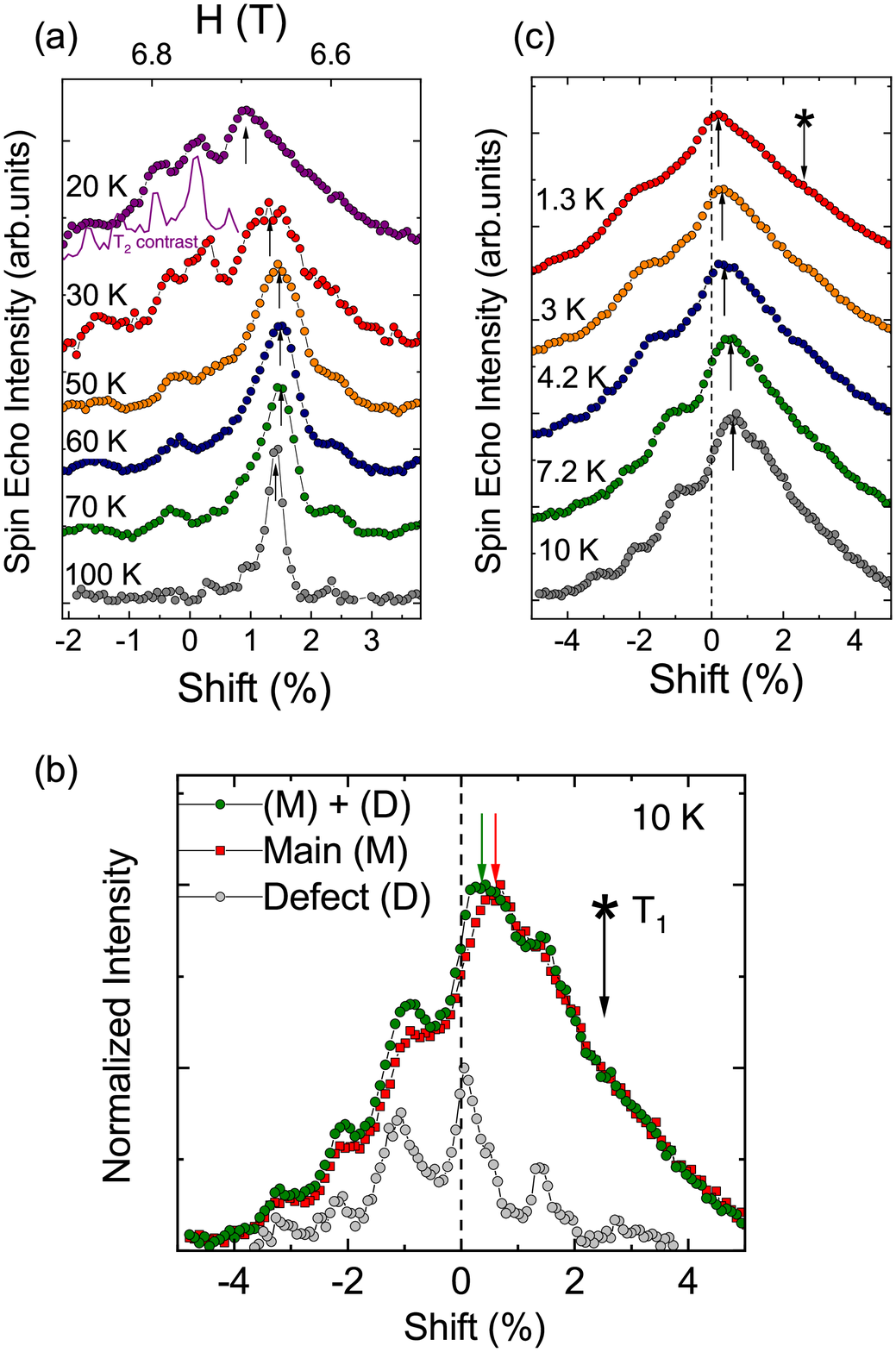}
\end{center}
Fig.~2:~{\bf Low-$T$ NMR spectra measured for $B\parallel a^*$ }\\


a) Bottom to top: evolution of the swept field spectra taken in an intermediate $T$-range. For $B\parallel a^*$, the spectrum resumes in a single line for O(1) site (fig.~1b) which dominates over other contributions, in particular the O(2) one which remains splitted. The shift of that main line can then easily be tracked (arrows) and goes through a maximum around 60 K $\sim J/3$ which cannot be observed in macroscopic susceptibility (see SI.~1). The spectra constantly broaden whereas a second set of well-resolved lines associated with defects surges at low temperatures which can be isolated using contrast experiments (purple line on the 20~K spectrum).\\
b) 10~K spectra as an example of the low-$T$, $T_2$ contrast procedure to reveal the defect contribution (grey). The contribution of the defect in the overall spectrum (green) can be nearly cancelled in the main spectrum (red) by saturation of the defect contribution ($T_1$ contrast described in the text). More examples at lower $T$ are given in SI.~3. The typical intensity contribution of the defect spectrum measured at the maximum is estimated to be of the order of 20\%. The arrows point to the maxima of the spectra from which the shifts are calculated. The (*)- black arrow points to the field at which the $T_1$ measurement was performed, away from the defect contribution.\\
c) Low-$T$ spectra obtained using the saturation procedure. The arrows monitor the shift which reveals the susceptibility of the kagome planes. $T_1$ measurements have been taken at the position of the (*) symbol in panels (b) and (c), that is far away from the defect spectrum.
\clearpage
\begin{center}
\includegraphics[width=1\textwidth]{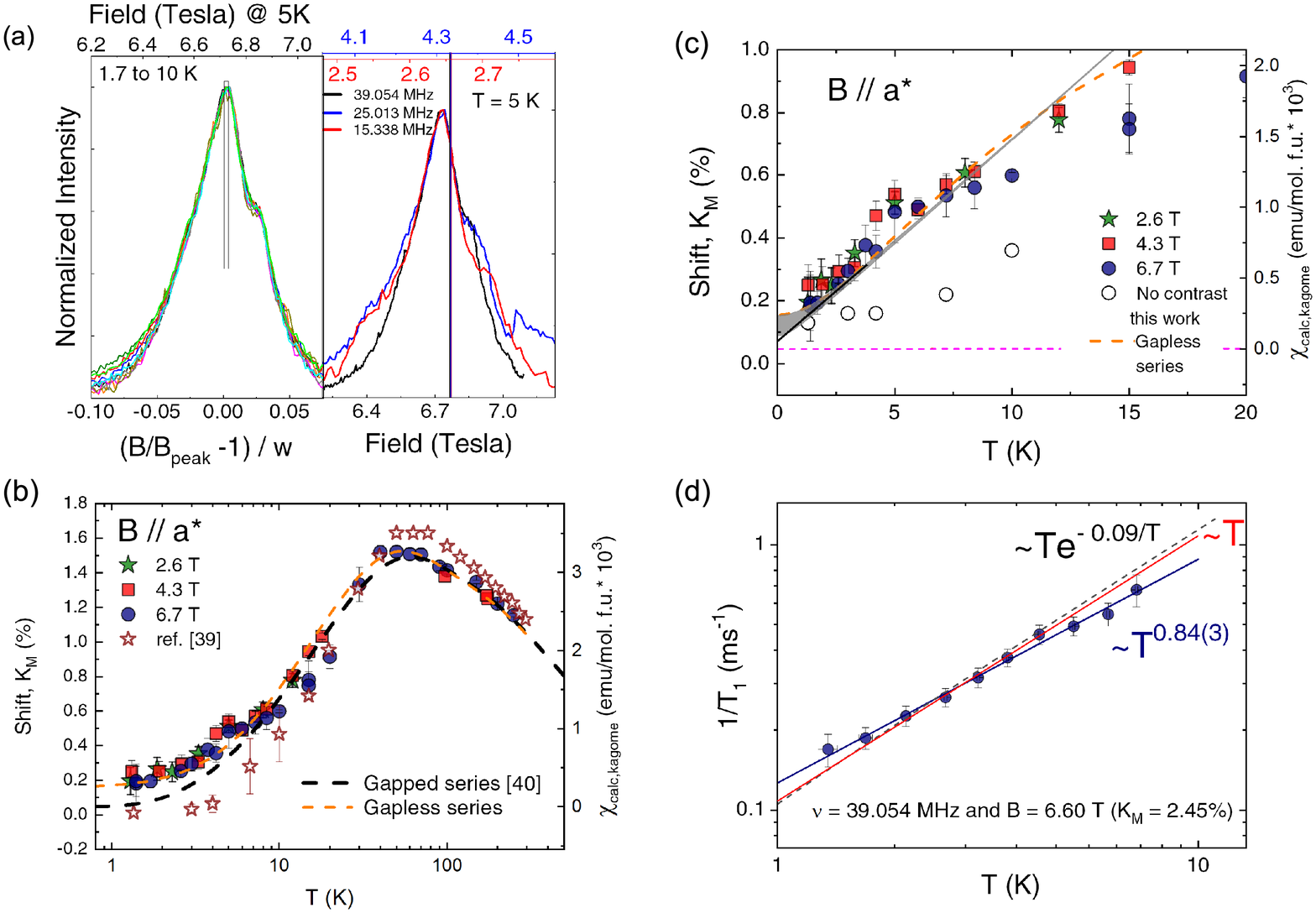}
\end{center}
Fig.~3:~{\bf Gapless ground state from shift and $T_1$ measurements}\\
 (a) Low-temperature spectra obtained by sweeping the field $B$ at a constant frequency $\nu$, corresponding to the reference field $\displaystyle \big( B_{\rm ref}=\frac{2\pi\nu}{^{17}\gamma}\big)$ with $^{17}\gamma = 2\pi\times 5.7718$~MHz/T. The shift of the main line, $K_{\rm M}$,  is obtained from the peak position in the spectrum at a field $B_{\rm peak}$, $\displaystyle \big( \frac{B_{\rm ref}}{B_{\rm peak}}=1+K_{\rm M}\big)$. Left panel:~The spectra are rescaled by a factor w proportional to their width and shifted horizontally to bring them into a perfect coincidence for an applied field $\sim$ 6.7~T and for temperatures ranging from 1.7 to 10~K. This enables an unprecedented accuracy on the determination of the relative variation of the NMR shift with temperature. The rectangle figures the typical error bar in the determination of the shift. Right panel: the same processing of data is performed at $T=5$~K, for various irradiation frequencies,  corresponding to fields ranging from 2.5 to 6.7~T.
  The shifts, measured through the distance to the reference fields (vertical bar), differ by only 0.03\% from the average value 0.52\%. This is within the $\sim 0.05\%$
  absolute error bar on each shift determined separately. We can safely conclude that the shift, therefore the susceptibility is field-independent.\\
(b)	Semi-log plot of the shifts taken for various reference fields and temperatures (log-scale). The small discrepancy with the data from ref.~\cite{Fu} at high-$T$ is
likely due to a slightly different orientation of the crystal. The large difference at low-$T$ originates from the refinement of our experimental method. The dashed lines
are from series obtained using the microcanonical approach described in the text with an origin for the spin shift at $K_{\rm chem}=0.046\%$. The orange line is sketched
assuming a $T\rightarrow 0$ linear specific heat and a finite $\chi(T=0) = 2.4\;10^{-4}$~emu/mol. f.u. obtained in the Dirac model under a 6.7~T field for
$v_F=5.9\times10^3$~m/s.\\
(c)	Same plot as (b) in a linear $T$-scale, with a zoom on low-$T$. The ultimate behavior of the kagome susceptibility is quasi-linear implying no gap; Our shift data can be either analyzed in the Dirac spin liquid framework (grey pattern with borders corresponding to 2.6 and 6.7 T) or the shift can be extrapolated linearly (black line) to a value either corresponding to or being larger than the chemical shift. In the latter scenario, at $T=0$ one would respectively deduce $\chi_{\rm kagome}=0$ or  $\chi_{\rm kagome}$ finite, less than 7\% of the maximum susceptibility for $T\sim J/3$ .\\
(d)	The spin-lattice relaxation rate of the main line is quasi-linear in $T$ (no gap). The blue line figures the best power law fit while the red one is for a $T$-linear variation. The dashed line displays the best gap scenario consistent with our data, $1/T_1\sim T\exp(-\Delta/T)$ with a maximum $\Delta \sim 5\;10^{-4}\;J = 90$~mK.
\newpage
\begin{center}
\centering
\includegraphics[width=1\textwidth]{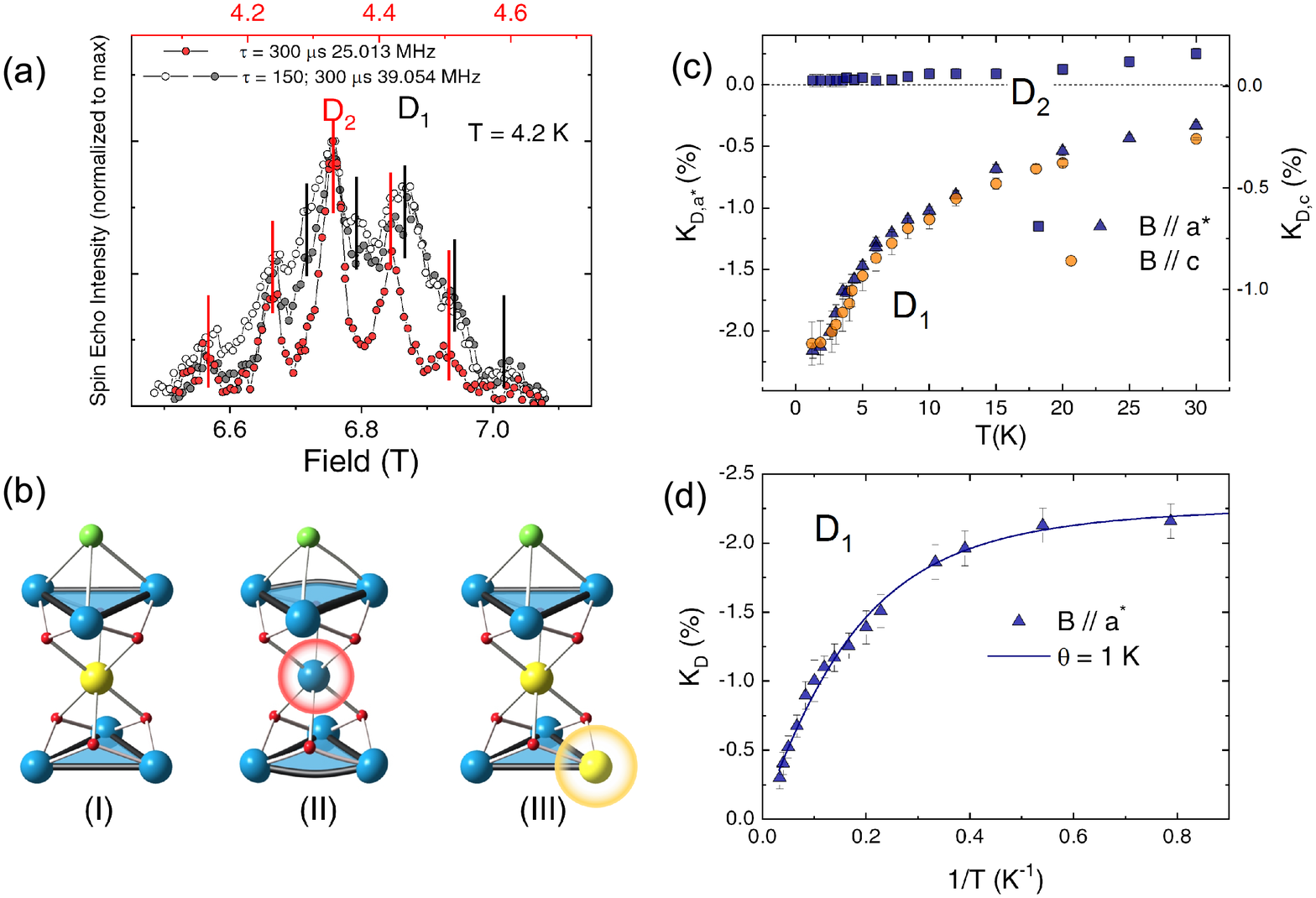}
\end{center}
Fig.~4:~{\bf Defect lines and local configurations}\\
(a) Low-$T$ $T_2$ contrast spectra enhancing the contribution of the defect over the main spectra. The comparison of such spectra obtained for different fields ranging from 2.6 to 6.7~T (two of them are shown here) enable to isolate two contributions from oxygen sites with a different defect environment. The vertical lines mark the 5 peaks of the negatively shifted (black, D$_1$) and the nearly unshifted (red, D$_2$) defect spectra.\\
(b) Structures around oxygens. Cu's are represented in blue, Zn's in yellow, Cl's in green and oxygens in red. (I) No {\it n.n} defect; (II) Cu$^{2+}$ replacing interlayer Zn$^{2+}$. Due to the Jahn-Teller character of the Cu$^{2+}$ ion, a local deformation is expected which might slightly impact the scheme of exchange interactions between Cu's in the kagome plane. The oxygen nucleus is still dominantly coupled to the Cu's of the kagome planes and weakly coupled to the Cu substituted on the Zn site. (III) Zn substitutes a « kagome » Cu, inducing then a magnetic vacancy in the kagome plane; the adjacent oxygens are coupled to only one Cu.\\
(c) $T$-plot of the shift for the two defects contributions D$_1$ and D$_2$ detected in our low-$T$ NMR experiments (see SI.~3); D$_1$ clearly mimics the low-$T$ macroscopic susceptibility which is dominated by the quasi-free contribution of Cu's substituted on the Zn site (II). D$_2$ has a negligible shift.\\
(d) Inverse $T$-plot of the most shifted (D$_1$) defect spectrum. The line is a fit to a Brillouin law with a small Weiss temperature issued from shared fits of the width at different fields $\theta=1(1)$~K (SI.~7).\\

\noindent{\bf Methods}\\
The single crystals were grown using the same method as in~\cite{Han-susc}.
In more details, 1.008~g of 4N ZnCl$_2$ (dried, when necessary, at 150$^\circ$C for 48 h) and 118~mg of 4N5 CuO (grinded) powders were first weighed inside a dry glovebox (1~ppm H$_2$O). While the latter was poured at the bottom of a cleansed quartz ampoule, the former was dissolved in exactly 2.3~mL of 30\% $^{17}$O-enriched water solution.
The ZnCl$_2$-rich and $^{17}$O-enriched solution was then poured in the ampoule already containing the CuO powder. The quartz ampoule was then sealed under a low static vacuum (0.7~atm), and slowly heated in an oven at 120$^\circ$C (24~h), and finally at 180$^\circ$C (48~h). The pale emerald green precipitate that forms at this stage, with no trace of unreacted black CuO, is the expected herbertsmithite phase, however poorly crystallized, as shown by routine x-ray powder diffraction. The crystal growth was then undertaken in a transparent 3 independent zones furnace, programmed in such a way that the growth temperature at the bottom of the temperature sink between the first two heating zones is 180$^\circ$C. 6 months of growth were necessary to achieve the crystal investigated in this work. The Cl, Cu and Zn amounts in the formed crystals was determined by EPMA/WDS coupled with EDS analysis. It was necessary to optimize counting times, to prepare fresh standards and to work on powders (by grinding single crystals) to obtain reliable and reproducible data by EPMA/WDS and EDS. While the Cl/Cu ratio always remains 3/2, the Cu/Zn one was found to be 3.3, which leads to an average $x$ value of 0.93 (with $x=4$/[1+Cu/Zn]). x-ray diffraction performed on powders sieved down to 63 microns was analyzed by means of Le Bail refinements. No other phase than the rhombohedral one (space group R-3m) was found and the room temperature lattice parameters extracted from the refinement procedure were $a=6.83791$~{\AA} and $c=14.0836$~{\AA} ($a=6.83109$~{\AA} and $c=14.04545$~{\AA} at 100~K). The crystals were also found to be highly faceted and the orientation of the facets was determined by single crystals XRD using a Laue camera in backscattering geometry. Only (-10-1) or (101) facets were found, which confirms the observations of [1], and can be explained by the fact that it corresponds to the highest lattice spacing obtained in this structure ($5.455$~{\AA}). 
The mass of the crystal used in our NMR experiments was $\sim 6$ mg.

The macroscopic magnetic properties were characterized using both MPMS Quantum Design SQUID magnetization measurements under fields ranging from 0.01 to 1~T, between 1.8 and 300~K and torque measurements under a 0.2~T field from 1.8 to 300~K.

Our NMR spectra at high-$T$ were obtained 
 by varying the frequency through all the transitions, then recombining the various Fourier transforms. Below 100~K, our spectra were obtained using field sweeps at a given frequency for typical frequencies 39.054; 25.013; 15.338~MHz used in most cases, corresponding to reference fields of 6.766; 4.334; 2.657~T. The delay $\tau$ used in the "$\pi /2 - \tau - \pi"$ spin-echo sequence was typically 14~$\mu$s for the spectra obtained at short times, mixing the defect and the intrinsic contribution. Delays $\tau$ of 150 up to 600 $\mu$s were used, depending on temperature, in order to isolate the defect contribution which has a longer transverse relaxation time, $T_2$ ($T_2$ contrast method). In order to single out the kagome contribution, we used a $\pi/2$ pulse applied 100 $\mu$s before the spin echo sequence to saturate the defect contribution which has a longer $T_1$ relaxation time ($T_1$ contrast method).

$T_1$ measurements were performed using a standard saturation recovery sequence at the main field under study $\sim 6.6$ T. We 
took our data on a part of the raw spectra where defect lines marginally contribute (fig.~2).

The chemical shift and the hyperfine constant used to derive Eq.~(1) were obtained from a linear fit of the shift versus kagome susceptibility plot. In order to avoid the defect contribution still present at high-$T$ at a level $\sim 10$\%, the intrinsic kagome susceptibility  was  calculated through series expansion and Pade approximants which prove to give reliable values for data between 200~K and 300~K. Prior to this, we indeed checked that a correct fit of the measured macroscopic susceptibility for both directions reported in \cite{Han-susc} was obtained by adding the $\sim 1/T$ contribution of defects to that calculated kagome susceptibility. The Land{\'e} $g$-factors for both the defect and the kagome contributions in the two directions of the applied field used in that calculations were obtained from ESR data published in~\cite{Zorko-ESR2}.

\noindent {\bf Data availability}\\
The dataset in our manuscript are available from the corresponding authors upon reasonable request.

\medskip

\noindent {\bf Acknowledgements}\\
This work was supported by the French Agence Nationale de la Recherche under Grants No. ANR-12-BS04-0021 'SPINLIQ' and No. ANR-18-CE30-0022-04 'LINK', and by Universit\'e Paris-Sud Grant MRM PMP. P.K. acknowledges support from the European Commission
through Marie Curie International Incoming Fellowship (PIIF-GA-2013-627322). We thank J. Quilliam and G. Simutis for a critical reading of the manuscript.
\medskip

\noindent {\bf Author contributions}\\
P.M. and F.B. conceived, designed and led the project. M.V. grew and characterized the single crystal. P.K, P.M., A.L. and Q.B carried out the NMR measurements and analysis. A.Z. carried out the ESR experiments. L.M. and B.B. performed the calculations of series. E.K, F.B. and P.M. supervised part of the experimental work and discussed the results.  P.M. wrote the manuscript with feedback from all the authors.
\medskip

\noindent {\bf Additional information}\\
The authors declare no competing financial interests. Correspondence and requests for materials should be addressed to P.M.
\medskip

\noindent {\bf Corresponding author}\\
Correspondence to Philippe Mendels philippe.mendels@u-psud.fr

\medskip

\begin{thebibliography}{99}
\bibitem{Knolle}Knolle, J. and Moessner, R. A Field Guide to Spin Liquids {\it Annual Review of Condensed Matter Physics} {\bf 10} 451-472 (2019).
\bibitem{Wen}Wen, X. G. Quantum orders and symmetric spin liquids {\it Phys. Rev. B} {\bf 65,} 165113 (2002).
\bibitem{Savary}Savary, L. \& Balents, L. Quantum spin liquids: a review. {\it Rep. Prog. Phys.} {\bf 80,} 016502 (2017).
\bibitem{Lacroix-Springer}{\it Introduction to Frustrated Magnetism} (eds Lacroix, C., Mendels, P. \& Mila, F.) (Springer, 2010).
\bibitem{Balents}Balents, L. Spin liquids in frustrated magnets. {\it Nature} {\bf 464,} 199-208 (2010).
\bibitem{Mendels-JPSJ}Mendels, P. \& Bert, F. Quantum kagome antiferromagnet $\mathrm{ZnCu_3(OH)_6Cl_2}$. {\it J. Phys. Soc. Jpn.} {\bf 79,} 011001 (2010).
\bibitem{Mendels-academy} Mendels, P. \& Bert, F.  Quantum kagome frustrated antiferromagnets: One route to qunatum spin liquids. {\it C.R. Physique} {\bf 17,} 455 (2016).
\bibitem{Zhou-KanodaRMP}Zhou, Y. Kanoda, K. \& Ng T.K. Quantum spin liquid states. {\it Rev. Mod. Phys.} {\bf 89,} 025003 (2017).
\bibitem{Shores}Shores, M. P., Nytko, E. A., Bartlett, B. M. \& Nocera, D. G. A structurally perfect S=1/2 kagom\'e antiferromagnet. {\it J. Am. Chem. Soc.} {\bf 127,} 13462-13463 (2005).
\bibitem{Jeschke} Jeschke, H.O., Pujol, F.S.  \& Roser Valenti, R. First-principles determination of Heisenberg Hamiltonian parameters for the spin-1/2 kagome antiferromagnet $\mathrm{ZnCu_3(OH)_6Cl_2}$. {\it Phys. Rev. B} {\bf 88,} 075106 (2013).
\bibitem{Mendels-musr}Mendels, P. {\it et al}. Quantum magnetism in the paratacamite family: Towards an ideal kagom\'e lattice. {\it Phys. Rev. Lett.} {\bf 98,} 077204 (2007).
\bibitem{Han-neutrons}Han, T.H. {\it et al}. Fractionalized excitations in the spin-liquid state of a kagome-lattice antiferromagnet.{\it Nature} {\bf 492}, 406-410 (2012).
\bibitem{Norman}Norman, M. R. Herbertsmithite and the search for the quantum spin liquid. {\it Rev. Mod. Phys.} {\bf 88,} 041002 (2016).
\bibitem{Barlowite}Han, T.H. {\it et al}. Barlowite: A Spin-1/2 Antiferromagnet with a Geometrically Perfect Kagome Motif.  {\it Phys. Rev. Lett.} {\bf 113,} 227203 (2014).
\bibitem{BarlowiteNMR} Feng, Z. et al. Gapped spin-1/2 spinon excitations in a new kagome quantum spin liquid compound Cu$_3$Zn(OH)$_6$FBr. {\it Chin.
Phys. Lett.} {\bf 34,} 077502 (2017).
\bibitem{Brochantite}Li, Y. Gapless quantum spin liquid in the S = 1/2 anisotropic kagome antiferromagnet ZnCu$_3$(OH)$_6$SO$_4$,
{\it New J. Phys.} {\bf 16}, 093011 (2014).
\bibitem{PuphalGa}Puphal, P. {\it et al}. Tuning of a Kagome Magnet: Insulating Ground State in Ga-Substituted $\mathrm{Cu_4(OH)_6Cl_2}$. {\it Phys. Status Solidi B}, 1800663 (2019).
\bibitem{PuphalY}Puphal, P. {\it et al}. Strong magnetic frustration in $\mathrm{Y_3Cu_9(OH)_{19}Cl_{18}}$: a distorted kagome antiferromagnet. {\it J. Mater. Chem. C} {\bf 5,} 2629 (2017).
\bibitem{ChineseY}Sun, W. {\it et al}{\it et al}. Perfect Kagom{\'e} lattices in $\mathrm{YCu_3(OH)_6Cl_3}$: a new candidate for the quantum spin liquid state. {\it J. Mater. Chem. C} {\bf 4,} 8772 (2016).
\bibitem{BartelemyY}Barth{\'e}lemy, Q. {\it et al}. Local study of the insulating quantum kagome antiferromagnets YCu$_3$(OH)$_6$O$_x$Cl$_{3-x}$ ($x$ = 0, 1/3). arXiv: 1904.04125v1.
\bibitem{McQueen}Kelly, Z.A., Gallagher, M. J. \& McQueen, T. M. Electron Doping a Kagome Spin Liquid. {\it Phys. Rev. X} {\bf 6,} 041007 (2016).
\bibitem{VOF}Clark, L. {\it et al}. Gapless Spin Liquid Ground State in the S=1/2 Vanadium Oxyfluoride Kagome Antiferromagnet $\mathrm{[NH_4]_2[C_7H_{14}N][V_7O_6F_{18}]}$. {\it Phys. Rev. Lett.} {\bf 110,} 207208 (2013).
\bibitem{Orain}Orain, J. C. {\it et al}. Nature of the Spin Liquid Ground State in a Breathing Kagome Compound Studied by NMR and Series Expansion. {\it Phys. Rev. Lett.} {\bf 118,} 237203 (2017).
\bibitem{Singh}Singh, R. R. P. \& Huse, D. A. Ground state of the spin-1/2 kagome-lattice Heisenberg antiferromagnet. {\it Phys. Rev. B} {\bf 76,} 180407(R) (2007).
\bibitem{Ran}Ran, Y., Hermele, M., Lee, P. A. \& Wen, X.-G. Projected-wave-function study of the spin-1/2 Heisenberg model on the kagome lattice. {\it Phys. Rev. Lett.} {\bf 98,} 117205 (2007); Hermele, M., Ran, Y.,  Lee, P. A. \& Wen, X.-G. Properties of an algebraic spin liquid on the kagome lattice Phys. Rev. B {\bf 77}, 224413 (2008); Ran, Y., Ko, W. -H., Lee, P. A. \& Wen, X.-G. Spontaneous Spin Ordering of a Dirac Spin Liquid in a Magnetic Field. {\it Phys. Rev. Lett.} {\bf 102,} 047205 (2009).
\bibitem{Lhuillier}Sindzingre, P. \& Lhuillier, C. Low-energy excitations of the kagome antiferromagnet and the spin-gap issue. {\it Europhysics Lett.} {\bf 88,} 27009 (2009).
\bibitem{Yan}Yan, S. Huse, D. A. \& White, S. R. Spin liquid ground state of the S=1/2 kagom\'e Heisenberg model. {\it Science} {\bf 332}, 1173-1176 (2011).
\bibitem{Depenbrock} Depenbrock, S. McCulloch,I. P. \& Schollwöck, U. Nature of the Spin-Liquid Ground State of the S=1/2 Heisenberg Model on the Kagome Lattice. {\it Phys. Rev. Lett.} {\bf 109,} 067201 (2012).
\bibitem{Iqbal}Iqbal, Y., Becca, F. Sorella, S. \& Poilblanc, D. Gapless spin-liquid phase in the kagome spin-1/2 Heisenberg antiferromagnet. {\it Phys. Rev. B} {\bf 87,} 060405(R) (2013).
\bibitem{Lauchli} L\"{a}uchli, A.M., Sudan, J. \& Moessner, R.  The S=1/2 Kagome Heisenberg Antiferromagnet Revisited. arXiv:1611.06990 (2016).
\bibitem{Hotta}Hotta, C. \& Asano, K. Magnetic susceptibility of quantum spin systems calculated by sine square deformation: One-dimensional, square lattice, and kagome lattice Heisenberg antiferromagnet. {\it Phys. Rev. B} {\bf 98,} 140405(R) (2018).
\bibitem{He}He, Y-C., Zaletel, M.P. Oshikawa, M. Pollmann, F. Signatures of dirac cones in a DMRG study of the Kagome Heisenberg model. {\it Phys Rev X} {\bf 7,} 031020 (2017).
\bibitem{Normand}Liao, H.J. {\it et al}. Gapless spin-liquid ground state in the S=1/2 kagome antiferromagnet. {\it Phys. Rev. Lett.} {\bf 118,} 137202 (2017).
\bibitem{deVries} M. A. de Vries, M. A. de. {\it et al}. Magnetic Ground State of an Experimental S = 1/2 Kagome Antiferromagnet. {\it Phys. Rev. Lett.} {\bf 100,} 157205 (2008).
\bibitem{Han}Han, T. H. {\it et al}. Thermodynamic Properties of the Quantum Spin Liquid Candidate $\mathrm{ZnCu_3(OH)_6Cl_2}$ in High Magnetic Fields. arXiv:1402.2693v1.
\bibitem{Bert}Bert, F. {\it et al}. Low temperature magnetization of the S=1/2 kagome antiferromagnet $\mathrm{ZnCu_3(OH)_6Cl_2}$. {\it Phys. Rev. B} {\bf 76,} 132411 (2007).
\bibitem{Han-susc} Han, T. H. {\it et al}. Synthesis and characterization of single crystals of the spin- 1/2 kagome-lattice antiferromagnets Zn$_x$Cu$_{4-x}$(OH)$_6$Cl$_2$. {\it Phys. Rev. B} {\bf 83,} 100402(R) (2011).
\bibitem{Olariu}Olariu, A. {\it et al}. $^{17}$O NMR study of the intrinsic magnetic susceptibility and spin dynamics of the quantum kagome antiferromagnet $\mathrm{ZnCu_3(OH)_6Cl_2}$. {\it Phys. Rev. Lett.} {\bf 100,} 087202 (2008).
\bibitem{Fu}Fu, M., Imai, T. Han,T.H. \& Lee, Y. S. Evidence for a Gapped Spin-Liquid Ground State in a Kagome Heisenberg Antiferromagnet. {\it Science} {\bf 350,} 655-658 (2015).
\bibitem{BernuLhuillier}Bernu, B. \& Lhuillier, C. Spin Susceptibility of Quantum Magnets from High to Low Temperatures. {\it Phys. Rev. Lett.} {\bf 114,} 057201 (2015).
\bibitem{Kawamura}Kawamura, H., Watanabe, K. \& Shimokawa, T. Quantum Spin-Liquid Behavior in the Spin-1/2 Random-Bond Heisenberg Antiferromagnet on the Kagome Lattice. {\it J. Phys. Soc. Jpn.} {\bf 83,} 103704 (2014).
\bibitem{graphene}Zhu, L. Wang, X.  Physics Letters A {\bf 380,} 2233 (2016)
\bibitem{Schnack}Schnack, J. Schulenburg,J. \& Richter, J. Magnetism of the N=42 kagome lattice antiferromagnet. {\it Phys. Rev. B} {\bf 98,} 094423 (2018).
\bibitem{Normand-bis}Lee, C. -Y.,  B. Normand, B. \& Kao, Y.J. Gapless spin liquid in the kagome Heisenberg antiferromagnet with Dzyaloshinskii-Moriya interactions. {\it Phys. Rev. B} {\bf 98,} 224414 (2018).
\bibitem{Zorko-ESR1}Zorko, A. {\it et al}. Dzyaloshinsky-Moriya anisotropy in the spin-1/2 kagome compound $\mathrm{ZnCu_3(OH)_6Cl_2}$. {\it Phys. Rev. Lett.} {\bf 101,} 026405 (2008).
\bibitem{ElShawish}El Shawish,S. Cepas, O. and Miyashita, S. Electron spin resonance in S=1/2 antiferromagnets at high temperature. {\it Phys. Rev. B} {\bf 81}, 224421 (2010)
.
\bibitem{Zorko-Vesi}Zorko, A. {\it et al}. Dzyaloshinsky-Moriya interaction in vesignieite: A route to freezing in a quantum kagome antiferromagnet. {\it Phys. Rev. B} {\bf 88,} 144419 (2013).
\bibitem{Tedoldi}Tedoldi, F., Santachiara, R. \& Horvatic, M. $^{89}$Y NMR Imaging of the Staggered Magnetization in the Doped Haldane Chain Y$_2$BaNi$_{1-x}$Mg$_x$O$_5$. {\it Phys. Rev. Lett.} {\bf 83,} 412 (1999).
\bibitem{Alloul-review}Alloul, H., Bobroff, J., Gabay, M. \& Hirschfeld, P. J. Defects in correlated metals and superconductors. {\it Rev. Mod. Phys.} {\bf 81,} 45-108 (2009).
\bibitem{Zorko-ESR2}Zorko, A. {\it et al}. Symmetry Reduction in the Quantum Kagome Antiferromagnet Herbertsmithite. {\it Phys. Rev. Lett.} {\bf 118,} 017202 (2017).
\bibitem{McQueen2} Freedman, D. E. {\it et al}. Site Specific X-ray Anomalous Dispersion of the Geometrically Frustrated Kagome Magnet, Herbertsmithite, $\mathrm{ZnCu_3(OH)_6Cl_2}$. {\it J. Am. Chem. Soc.} {\bf 132,} 16185 (2010).
\bibitem{Rousochatzakis}Rousochatzakis, I., Manmana, S. R., Lauchli, A. M., Normand, B. \& Mila, F. Dzyaloshinskii-Moriya anistropy and nonmagnetic impurities in the $S =\frac{1}{2}$ kagome system ZnCu$_3$(OH)$_6$Cl$_2$. {\it Phys. Rev. B} {\bf 79,} 214415 (2009).
\bibitem{Kondo-Brochantite}Gomil\v{s}ek. M {\it et al}. Kondo screening in a charge-insulating spinon metal. {\it Nat. Phys.}{\bf 15,} 754 (2019). \end{thebibliography}
\end{document}